\documentclass[10pt]{amsart}
\usepackage{enumerate}
\usepackage{graphicx,graphics}
\usepackage{amsfonts}
\usepackage{amssymb}
\usepackage{amsthm}
\usepackage{amsmath}
\usepackage{mathrsfs}
\input{xy}
\xyoption{all}

\newtheorem{theorem}{Theorem}[section]
\newtheorem{lemma}[theorem]{Lemma}

\newtheorem{remark}[theorem]{Remark}
\newtheorem{corollary}[theorem]{Corollary}

\newtheorem{proposition}[theorem]{Proposition}

\numberwithin{equation}{section}

\title[Dreaming machine learning]{Dreaming machine learning: Lipschitz extensions for reinforcement learning on financial markets}

\author{J.M. Calabuig, H. Falciani and E.A. S\'anchez-P\'erez}

\address{Instituto Universitario de Matem\'atica Pura y Aplicada. Universitat Polit\`ecnica de Val\`encia. 46022 Valencia. Spain\\
jmcalabu@mat.upv.es, herfal@upvnet.upv.es, easancpe@mat.upv.es }



\thanks{The third  author gratefully acknowledge the support of the  Ministerio de
Econom\'{\i}a y Competitividad (Spain) and FEDER under grant   MTM2016-77054-C2-1-P }

\begin{document}
\maketitle

\begin{abstract}
We consider a quasi-metric topological structure for the construction of a new reinforcement learning model in the framework of financial markets. It is based on a Lipschitz type extension of reward functions defined in metric spaces. Specifically, the McShane and Whitney extensions are considered for a reward function which is defined by the total evaluation of the benefits produced by the investment decision at a given time. We define the metric as a linear combination of a Euclidean distance and an angular metric component. All information about the evolution of the system from the beginning of the time interval is used to support the extension of the reward function, but in addition this data set is enriched by adding some artificially produced states.  Thus, the main novelty of our method is the way we produce more states ---which we call ``dreams"--- to enrich learning. Using some known states of the dynamical system that represents the evolution of the financial market, we use our technique to simulate new states by interpolating real states and introducing some random variables. These new states are used to feed a learning algorithm designed to improve the investment strategy by following a typical reinforcement learning scheme. 

\end{abstract}



\section{Introduction}

The theory of Lipschitz functions in metric spaces is a theoretical tool that has often been considered since the beginning  of  machine learning. Indeed, several theoretical aspects on Lipschitz extension of maps which can be interpreted as foundations of reinforcement learning procedures were published in some early papers many years ago. Just as an example, in the 1967 paper \cite{aron}
the reader can find some applications of the McShane and Whitney extensions---which are called the lower and upper functions in this paper---to issues that can be identified now as machine learning problems. In addition, the reader can find in \cite{milman} some results on so called absolutely minimal extensions, which can also be considered as mathematical foundations of extension procedures associated with machine learning methods. Moreover, there are some explicit applications of Lipschitz functions to machine learning in  \cite{asadi,dywx,lux} and the references therein. 
However, the main relation of Lipschitz functions with the mathematical framework of machine learning 
is associated with the notion of Lipschitz continuity, which allows to control the regularity and smoothness of the functions involved, as can be seen for example in \cite{asadi} and \cite[Sec.2]{xzlq}. A full explanation of why these requirements are necessary in several machine learning techniques can be found in \cite{ca}; the reason is the same as that which makes Lipschitz's condition so relevant in control theory (see for example \cite{dpddv,gwlt,lzys}).

Our paper presents a new technique that uses the theory of Lipschitz functions in a different way. We focus our attention on the McShane-Whitney type extension of the Lipschitz functions as the main tool for predicting the value of the reward function in the next step of a dynamical system. There are not many papers in which Lipschitz extensions are used in this way.  A remarkable exception is provided by  \cite{dywx}---see also the references in Sections II and III of this paper---, which explicitly  mention the Lipschitz extension as a tool for solving classification problems, in the context of metric learning. We use the same fundamental extension results but in a different context. We construct  a typical reinforcement learning procedure, in which the forecast of the reward function is performed as a McShane-Whitney extension of its previous values. The underlying metric space is defined by the set of all the previous states---represented as vectors of a finite dimensional space---, for which the reward function is known, and another set that is artificially generated, whose elements we call ``dreams". The distance used to endow the set of states---the union of previous states and dreams---with a  metric structure is constructed as a weighted combination of a standard norm distance and an angular term. It will be explained in Section \ref{S3}, and
we have to point out that it is rather atypical. Often, the metric  structure underlying the extension of the Lipschitz maps is the usual finite dimensional space $\mathbb R^n$ with the Euclidean norm, or some classic modifications of this metric considering non-canonical scalar products acting on $\mathbb R^n$. Information on other related metric structures in which Lipschitz extensions of reward functions have been considered can be found, for example, in \cite{kyng,rao}, where metric graphs are studied.


Therefore, the goal of this paper is to show a new mathematical environment based on the McShane-Whitney reward function extensions for the development of a new reinforcement learning procedure.
However our ideas, which can be applied in much more general contexts, will focus on the rather specific issue of designing expert systems for financial market analysis.
The reason is that  financial time series  is one of the main challenges both in the time series theory  and the machine learning developments. Several approaches have been proposed in the last decades to predict financial time series as a fundamental step for constructing decision-making support systems. The reader can find some general explanation about the state of the art in \cite{cbsno,gmbc,sjz}.   The traditional framework is given by statistical/probabilistic  methods as stochastic Markov processes, that are associated to linearity assumptions on the processes that generate the time series. Although they have been successful in some cases, they have proved to be insufficient, mainly due to the great complexity of the financial time series, which makes it necessary to apply other methods that are adapted to the nature of the signals 
  (see for example \cite{lcks,ysx}).

Machine learning methods have also  shown to be useful for the analysis of financial time series. 
 Neural network based developments are probably the most popular \cite{ljyw,mdm,mme,tv}, often using feed forward schemes to introduce dynamically updated information \cite{mhyw}. Also, the advanced derivatives known as deep learning techniques \cite{chp,fk,s}, recurrent neural network \cite{mrm} and convolutional neural networks \cite{so}, sometimes mixed with other techniques such as probabilistic tools \cite{kdh,t}, have been widely used.
A relevant reason for the widespread use of neural networks is given by the fact that these structures are capable of handling data with components characterized by lack of smoothness (and therefore with bad polynomial approximations), discontinuity and (local) nonlinearity. Neural networks are self-adaptive methods, which are directly based on data and can process nonlinear time series behaviors without any statistical assumption on the data \cite{lw,llc}.  Thus, these techniques have generally proved to be better adapted to highly nonlinear processes and useful for both short and long term prediction (\cite{pstk}).  This is the reason why  we use this method to compare with our procedure in the present paper.

An alternative to artificial neural networks are support vector machines, a supervised learning statistical intelligent technique for data analysis and pattern recognition  that is suitable for highly nonlinear problems, and  can be used for both classification and regression \cite{zhq}. They are a good option for time series forecasting, and have been successfully applied in financial contexts \cite{ch,drfp,yhl}, often mixed with other methods \cite{dp,whw}.

The technique proposed in this paper does not follow any of the methods explained above. It is a reinforcement learning procedure that is not based on any of the approaches explained, but is closely related to some of the classical time series forecasting methods. Reinforcement learning techniques and the related  evolutionary computation \cite{zhxzsl}  concerns a great variety of methods, that are mainly characterized by the fact that the so called (software) agents  take actions in a dynamic context in such a way that they have to maximize some notion of cumulative reward, that is defined to be a key issue of the model \cite{sb}. 
They have been successfully applied in a wide variety of dynamic financial problems \cite{bc,lde,pc}.
It is the third classic option for dealing with problems such as those we are studying in this paper, in addition to supervised learning and unsupervised learning to which the techniques explained above belong \cite{mm}. Several methods are considered in this cathegory, as Q-learning \cite{dzl,lpjlh,psc}, recurrent \cite{ay,dzl} and adaptive \cite{dl,l} reinforcement learning techniques, often introducing tools from other contexts \cite{b,b2,mr}.  So called deep reinforcement learning is a recent theoretical context in which classical reinforcement learning techniques are used along with some other methods, mainly from later developments based on neural networks and other techniques \cite{dbkrd,jk,jl,zm}.

We  have to mention that, due to the high complexity of the problem of forecasting financial series, other techniques have also been used which are not related to the core of the methods we have explained. For example, a great effort has been made to add to these procedures arguments and tools from text mining \cite{nawn,whw0}, sentiment analysis \cite{kl,lxcwd} and semantic analysis \cite{bmz,cbsno,lm}.

Concerning related work on mathematical economy and models for financial markets, we
develop our method in a rather classical framework.
  The definition of our reward function begins with a relationship of duality similar to that of the
commodity-prize duality that is at the core of market models based on functional analytic tools (see for example \cite[Ch.8]{ali}).
Although our method refers to some probabilistic tools, we do not consider our learning method as based on stochastic arguments. However, philosophically we may refer to some links with  stochastic market modeling ---concretely to the so called continuous-time market model, see for example \cite[Ch.2]{korn}---, since the decision on the following step is given exactly in the previous one, based on a predictive reward function  in our case.

For clarity in the explanation of our technique, we will focus our presentation on  particular problems related to the dynamics of stock markets. As we said, \textit{our technique is based on significantly extending the reward function by creating new simulated situations to provide an improved tool for decision making in financial markets
based on Lipschitz preserving extensions of the reward.}  The calculations are simple, as the extension formulas are simple, so the technique could be applied when dealing with a large amount of data.

Our ideas will be presented in five sections. After this introductory section, we will explain
the topological foundations on the metric representation spaces that will be used in the preliminary Section \ref{S2}. In Section \ref{S3} we will describe the general facts for the definition of our procedure ---mainly of mathematical nature---, and the fundamental scheme of our algorithm. The construction of the associated models  will be presented in a very concrete way in Section \ref{S3bis}, in which two complete examples are developed. The results of both will be presented in Section \ref{S4}, along with some comparisons between different situations and an alternative procedure based on neural networks. The document ends with some conclusions in Section \ref{S5}.

We should note that the objective of the present paper is  theoretical in nature, although  very explicit examples are given. We do not intend to give  an efficient algorithm for computing the mathematical elements that appear in the model in order to provide a concrete and effective tool:  we are interested in explaining the fundamentals of our method instead.


\vspace{0.5cm}

\section{Preliminaries and topological tools} \label{S2}

Our arguments bring together ideas from abstract topology on quasi-pseudo-metric spaces and Lipschitz maps, and practical computational tools for
extending Lipschitz functions on metric vector spaces in which the distance
is \textit{not} given by a standard norm coming from an inner product. In fact, our metric is not one of the classical distances used in machine learning (see for example the comments is Section I and Section II in  \cite{jia}). We use  the  McShane and the Whitney extensions for Lipschitz maps in a special way in order to extend some
reward functions defined by a novel  design. The process of introduction of ``dreams"
to increase the size of the training set needs also some topological tools based on average values computed on equivalence classes constructed by a specific metric similarity method.
Although our mathematical approach is as far as we know new, the reader can find some related ideas in \cite{asadi,driesetal}.

A quasi-pseudo-metric  on a set $M$ is a function $d: M \times M \to \mathbb R^+$ ---the  set of non-negative real numbers--- such that
\begin{enumerate}
\item $d(a,b)=0$ if  $a=b$, and
\item $d(a,b) \leq d(a,c)+d(c,b)$
\end{enumerate}
for $a,b,c \in M$. A topology is defined by such a function $d$: the open balls define the  basis of neighborhoods. For $\varepsilon >0$, we define the ball of radius $\varepsilon$ and center in $a \in M$ as
$$
B_\varepsilon(a):= \Big\{ b \in M: d(a,b) < \varepsilon \Big\}.
$$
$(M,d)$ is called a quasi-pseudo-metric space.
We will work in this paper mainly with pseudo-metrics, that is,  $d(a,b)=d(b,a)$ for all $a,b \in M$, or metrics, that in addition satisfy
that  $d(a,b)=0$ if  and only if $a=b.$ 
In this case, the topology  defined by $d$ satisfies the  Hausdorff separation axiom.
However, we prefer to present some of our ideas in this more general context, since the basic elements of our technique can be easily extrapolated to the more general  quasi-pseudo-metric case. This fact is relevant, since asymmetry in the definition of metric notions (quasi-metric case) could be crucial for the modeling of dynamical processes, in which the dependence on the time variable changes the concepts related to distance.
As usual, we will use both the words metric and distance as synonyms.
We will use also classical notation for distances from a point to a set: if $d$ is a (pseudo-)metric in  a set $M$, $a \in M$ and $B \subset M,$ we will write $d(a,B)$ for the distance from $a$ to $B$, that is
$d(a,B)= \inf_{b \in B} \,d(a,b).$

Let us mention that another generalization of the notion of distance, the so called semimetrics, have been introduced recently in the context of  machine learning (see \cite{gkn}). These functions fail in the triangle inequality, while in general  quasi-pseudo-metrics fail in symmetry. The difference is relevant, since triangle inequality  is necessary for the formulas of McShane or Whitney to work as extensions,  which might not happen if a semimetric is used.

Let us recall now some definitions regarding functions. Let $(M,d)$  be a metric space. A  function $f: M \to \mathbb R$ is a Lipschitz function if
there is a positive constant $K$ such that
\begin{equation} 
| f(a)-f(b) |\le K \, d(a,b), \quad a,b \in M.
\end{equation} 
The infimum of such constants as $K$ is called the Lipschitz constant of $f$. We refer to \cite{cob} for a new complete explanation on this topic; all  general results on Lipschitz functions that are needed can be found there.

Regarding extensions of Lipschitz maps, recall that the classical McShane-Whitney theorem states that if
$(M_0,d)$  is a subspace  of a metric space $(M,d)$ and $T : M_0 \to \mathbb  R$ is a Lipschitz function with Lipschitz constant $K$, there always exists a Lipschitz function $T^M: M \to \mathbb R$  extending $T$ and with the same Lipschitz constant. There are also known  extensions of this result  to the general setting of real-valued semi-Lipschitz functions acting in quasi-pseudo-metric spaces; see for example \cite{aron,Mc,mustata,mustata2,romsan} and the references therein. The function
\begin{equation} 
{T^M}(a):=\sup_{b\in M_0}\{T(b)-K\,d(a,b)\}, \quad a \in M,
\end{equation} 
provides such an extension; it is sometimes called the McShane extension. We will use it for giving a constructive tool for our approximation. The Whitney formula, given by
\begin{equation} 
 T^W(a) :=\inf_{b\in M_0}\{T(b)+K\,d(a,b)\}, \quad a \in M,
\end{equation} 
provides also an extension. We will use the first one in this paper, although some results are also true when using the second, as will be explained. The reader can find more recent technical information directly related with our ideas in \cite{asadi,lux} and the references therein.
Concretely, some applied tools associated to   Lipschitz extensions of functions for machine learning can be found in \cite{gott,kyng,lux}. General explanations about applications of mathematical analysis in Machine Learning can be found in \cite{simovici}; in particular, basic definitions, examples and results on Lipschitz maps can be found in Section 5.10 of this book and in \cite{cob}.

We will use standard notation; we write  $\| \cdot\|_1$, $\| \cdot\|_2$ and $\| \cdot\|_\infty$ for the  $\ell^1,$ the $\ell^2,$ and  the $\ell^\infty$ norms respectively,  that will be called the $1$-norm, the $2$-norm (or the Eclidean norm), and the $\infty$-norm, as usual. If $X$ is a normed space, we denote by $B_X$ and $S_X$ the closed unit ball and the unit sphere of $X$, respectively.

\vspace{0.5cm}

\section{Metric spaces of states and Lipschitz maps: an algorithm  for machine learning} \label{S3}

We will model the set of  strategies to be applied in a financial market ---a dynamical system--- as a metric space of finite sequences of $n$ items ---states of the system---, where $n$ is the number of times that a change of state (purchase/sale event) could occur in the market.
We will consider also a reward function, that is supposed to be known for a certain subset of strategies ---initial ``training set"---. Using the well-known theoretical techniques of extension of Lipschitz functions on metric spaces that we have mentioned in the introductory section, we will construct the necessary tools for computing improved reward functions for bigger sets of strategies by means of the search of ``similarities" among different pieces of these items. This will be used to feed the algorithm for creating new situations ---``dreams"--- that will allow to increase the
efficiency of the process by increasing the size of the training set.
The final result will be the definition of a typical reinforcement learning method.


Consider a  subset  $M_0$ of vectors of the finite dimensional real linear space $\mathbb R^n$ not containing the $0$.  Let us write $M= \mathbb R^n \setminus \{0\}.$ We start by defining an adequate metric on $M$. As the reader will see, the difference of our technique with other methods of reinforcement learning begins at this point. The main reason is that our choice does not allow to
 define the distance by means of a norm in $\mathbb R^n$.
We mix the angular pseudo-distance ---geodesic distance--- and the Euclidean norm in this space. Thus, since the cosine of the angle among elements $s_1$ and $s_2$ in $M$ is given by
\begin{equation} 
Cos(s_1,s_2)= \frac{ s_1 \cdot s_2}{ \|s_1 \| \, \|s_2\|}, \quad s_1,s_2 \in M,
\end{equation} 
we define a distance by mixing this angle
\begin{equation} 
\Theta(s_1,s_2)= \frac{1}{ \pi} Arc Cos \Big(  \frac{ s_1 \cdot s_2}{ \|s_1 \| \, \|s_2\|} \Big),
\end{equation} 
and a Euclidean component
\begin{equation} 
E(s_1,s_2)= \| s_1- s_2\|_2 = \sqrt{\sum_{k=1}^n \big|s_{1,k}- s_{2,k} \big|^2 ,}
\end{equation} 
where $s_1=(s_{1,1},...,s_{1,n})$ and $s_2=(s_{2,1},...,s_{2,n}).$
This Euclidean term can be substituted by  any other norm in $\mathbb R^n$.
For each $\epsilon \ge 0,$ we define now the function
\begin{equation} \label{epdistance} 
d_\epsilon(s_1,s_2)= \Theta(s_1,s_2) + \epsilon  E(s_1,s_2), \quad s_1, \, s_2 \in M,
\end{equation}
that will become the general formula for the distance we want to use in our model. As usual, we use the
same symbol $d_\epsilon$ when it is restricted to any subset of $M.$ With this definition, we try to obtain a balance between the ``metric part" and the ``angular part" when comparing vectors representing states/actions. The metric part gives an estimate of the difference in vector sizes---which has a clear meaning in the model representing the differences in the ``volume" of investments---while the angular part gives an idea of the direction in which the market moves. The parameter $\epsilon$  allows us to modulate the weight we want to give to each term.

\begin{lemma} \label{lema}
Let $\epsilon >0$. With the definitions given above, the following statements hold.
\begin{itemize}
\item[(i)]
The function $d_\epsilon$ is a pseudo-metric on $M$ for every $\epsilon \ge 0.$ Moreover, it is a metric on $M$
if and only if $\epsilon >0.$

\item[(ii)] For every $\epsilon >0$, the metric space $d_\epsilon$ is (topologically) equivalent to  $E.$

\item[(iii)] Let $\epsilon >0$ and $S_0 \subset \mathbb R^n$ a set that includes an open segment containing $0$. Then, for any
extension $d_\epsilon^*$ of $d_\epsilon$ to $S_0,$ the metrics $d_\epsilon^*$ and $E$ are not equivalent on $S_0$.

\end{itemize}
\end{lemma}
\textit{Proof. 
(i) Note first that $\Theta$ is well-defined on $M.$
 The triangle inequality and the symmetry are satisfied by both the functions $\Theta$ and $E$. Indeed, it is known that $\Theta$ is a metric on the Euclidean unit sphere, and so if
$s_1,s_2,s_3 \in M$,
$$
\Theta(s_1,s_2) = \Theta(\frac{s_1}{\|s_1\|},\frac{s_2}{\|s_2\|})
$$
$$
\le
\Theta(\frac{s_1}{\|s_1\|},\frac{s_3}{\|s_3\|}) + \Theta(\frac{s_3}{\|s_3\|},\frac{s_2}{\|s_2\|})
=
\Theta(s_1,s_3) + \Theta(s_3,s_2).
$$
Moreover, any  linear combination with non-negative coefficients of $\Theta$ and $E$ is a pseudo-metric. Also,
if $\epsilon >0$ then $d(s_1,s_2)= \Theta(s_1,s_2) + \epsilon  E(s_1,s_2)=0$ implies $E(s_1,s_2)=0$, and so $s_1=s_2$. The converse is obvious too.}

\textit{
(ii) Take an element $s \in M$ and an open ball $B_{d_\epsilon,r}(s)$
of radius $r>0$ for the metric $d_\epsilon$. Take the elements $s' \in M$
in this set satisfying that $\Theta(s,s') <r/2$ and $E(s,s') <r/(2 \epsilon),$ and note that all of them are
in $B_{d_\epsilon,r}(s).$ Then, since $s \ne 0,$ by the continuity of $\Theta$ with respect to the Euclidean metric $E$
we can find a ball of radius $r' >0$  such that
$$
B_{E,r'}(s) \subset \{ s' \in M: \Theta(s,s') < r/2 \}.
$$
Thus, taking $r'' = \min \{r/(2 \epsilon) ,r'\}$ we get that $B_{E,r''}(s) \subseteq B_{d_\epsilon,r}(s).$
 The obvious inequality
$$
E(s_1,s_2)=\|s_1-s_2\|_2 \le \frac{1}{\epsilon} d_\epsilon(s_1,s_2), \quad s_1, s_2 \in M,
$$
gives the converse relation needed for the equivalence.
}

\textit{
(iii) Consider without loss of generality the vectors $b=(\alpha,0,0,0,...), \, -b=(-\alpha,0,0,0,...) \in M,$
for some $\alpha >0$. It is enough to notice that we can construct a sequence converging to $0$ with respect to $E$ and which does not converge for $d_\epsilon^*.$ Indeed,
$$
\lim_{0<\alpha \to 0} \|b-(-b)\|_2 = \lim_{0<\alpha \to 0} 2 \alpha = 0,
$$
but
$$
\lim_{0<\alpha \to 0} d_\epsilon(b,-b)=
\lim_{0<\alpha \to 0}  ArcCos \big(\frac{b \cdot (-b)}{\|b\| \, \|-b\|} \big)
+ \lim_{0<\alpha \to 0} \epsilon \|b- (-b)\|_2= 1.
$$
Thus, both metrics cannot be equivalent.
}

\vspace{0.3cm}

Of course, Lemma \ref{lema} can be automatically stated if we change the Euclidean norm by any other norm on $\mathbb R^n$, since all norms are equivalent on finite dimensional spaces. The metric $d_\epsilon$ is
defined to indicate the Euclidean distance among states $s_1$ and $s_2$ but also the trend that they represent: indeed, in terms of the financial model we are constructing, if two vectors have small size ---in fact as small as we want---, but they represent opposite trends in the market, the distance among them is always bigger or equal than $1.$ The relative weight of $\Theta$ and $E$ in the definition of $d_\epsilon$
is modulated by the parameter $\epsilon.$ 


We will define a reward function acting in $M_0$ that will be given, as a primary formula, by a duality relation among the elements $s \in M_0 \subset \mathbb R^n$ and vectors acting on these elements. We will call these vectors \textit{actions}, and they will be represented by vectors of (a multiple of) the unit sphere  of the space $(\mathbb R^n, \| \cdot\|)$ for any norm $\| \cdot \|.$ We will write $\mathcal A$ for the set of all actions to be considered in a problem.

We will define the reward $R: M_0 \to \mathbb R$ for a state $s$ as a function ---a maximum, or a mean--- of  actions that operates on $s$ as
\begin{equation} 
R_0(s) = s \cdot a, \quad s \in M_0, \,\,\,\, a \in B_s,
\end{equation} 
where $B_s$ is an $s$-dependent set defined using a mix among some experience on the system and a random procedure. The final function will be called $R$, the real function to be extended with the McShane formula for getting an estimate $R^M$ of the reward  acting in all the space $M$.

In any case, as we will see in the rest of the paper, it is always possible to write $R(s)$ as $s \cdot a_s$ for a given action $a_s$ belonging to the selected set of actions $\mathcal A$  for the elements $s$ of $M_0$. However, this representation formula cannot be obtained in general for all  the extended values $R^M(s^*),$
 $s^* \in M \setminus M_0,$ although some useful bounds can be obtained. Let us analyze this representation of $R$ and the associated bounds for the extension $R^M$ in what follows. Next example
shows that the extension $R^M$ cannot be written as the scalar product of  the state $s^*$ and an action
$a \in \mathcal A.$




\vspace{3mm}

\textit{Example.} 
Fix $\epsilon >0$.
Consider a market with two products  ($n=2$) and just two states.  Consider the set $M_0=\{(1,0),(2,0)\}.$ Both vectors represent increasing states of the market.
Consider the reward function given for both states  by the  actions $a_1=(50,50)$ and $a_2=(0,100),$
that define the set $\mathcal A.$ The
$1$-norm multiplied by $1/100$ is considered, that is,  both actions are elements of the set $100 \times S_{\ell^1}.$
 That is,
$R((1,0)):=(1,0) \cdot a_1= 50 $ and $R((2,0)):= (2,0) \cdot a_2=0.$
Note that $d_\epsilon((1,0),(2,0))=\epsilon.$
The Lipschitz constant $K$ is given by
$$
K=|0-50|/d_\epsilon ((1,0),(2,0)) = 50/\epsilon.
$$

Therefore, the McShane extension of $R$ is given by
$$
R^M((x,y)) := \max \big\{\, 50- (50/\epsilon) d_\epsilon((x,y), \, (1,0)), 0 - (50/\epsilon)  d_\epsilon((x,y),(2,0)) \, \big\}.
$$
for any possible state $(x,y) \in \mathbb R^2   \setminus \{0\}.$ Take now $(x,y)=(-1,0),$ and note that
$$
d_\epsilon((1,0),(-1,0))= 1+ 2 \epsilon \quad and  \quad d_\epsilon((2,0),(-1,0))= 1+ 3 \epsilon.
$$
 Then we have
$$
R^M((-1,0))= \max \{50- (50/\epsilon) d_\epsilon((-1,0),(1,0)), 0 - (50/\epsilon)  d_\epsilon((-1,0),(2,0))\}
$$
$$
= \max\{ 50-\frac{50}{\epsilon} \cdot (1+ 2 \epsilon), 0-\frac{50}{\epsilon} \cdot (1+ 3 \epsilon)\}.
$$
Take now $\epsilon = 1/2.$ Then
$
R^M((-1,0)) = \max \{-150, -250 \} = -150.
$
Since all the actions in $\mathcal A$ belong to the ball of radius $100$ of $\ell^1$, we cannot write
$
R^M((-1,0))= (-1,0) \cdot a
$
for any $a \in \mathcal A.$

\vspace{5mm}

In the rest of this section we show some boundedness results that compensate the lack of representation 
of the extension $R^M$ as a scalar product. Although this easy representation is not always possible, at least we can control the difference of the extension and the proposed formula using the scalar product. 
A relevant situation is given when the set of actions $\mathcal A$ is defined by multiples of partitions of the unity,  that is, sets of  vectors of the unit sphere of $\ell^1$ having all the coordinates bigger or equal than $0$. We will write  $S^{n,+}_{\ell^1}$ for this set, and we will consider the factor $100.$ This models the standard problem of distributing a fixed amount of money among a given set of products, depending on market forecasts.
 Thus, we will consider in this section the set $\mathcal A$ as a subset of $100 \, \times\, S^{n,+}_{\ell^1}$, in order to work with bets given as $\%$. However, the reader will notice that the same arguments and results can be easily adapted for the case of any other bounded set with respect to any norm.
Concretely, we consider the following  set of actions  over $M_0,$  
$$
\mathcal A_{M_0, R}=\{ a \in 100 \, \times \, S^{n,+}_{\ell^1}: a= a_s \,\textit{for some $s \in M_0$ such that} \,\, R(a)= s \cdot a_s \}.
$$


\vspace{0.3cm}

\begin{proposition} \label{propext}
Let $M_0 \subset M$ be a compact subset of  $(\mathbb R^n \setminus \{0\},\|\cdot\|_2)$. Consider a function $R:M_0 \to \mathbb R$ such that for each $s \in M_0$ there is a functional $a_s \in \mathcal A_{M_0, R} \subset 100 \, \times \, S^{n,+}_{\ell^1}$ such that
$$
R(s):= s \cdot a_s, \quad s \in M_0.
$$
Then for each $s^* \in M$ there is a functional $a_{s^*} \in \mathcal A_{M_0, R}$ such that
$$
|R^M(s^*)- s^* \cdot a_{s^*} | \le
\min_{s \in M_0} \Big( 100 \, \|s-s^*\|_\infty + K \Theta(s,s^*)+ \epsilon K E(s,s^*) \big) \Big).
$$
\end{proposition}
\textit{Proof.
Fix $s^* \in M.$
First note that, since $R^M$ is a Lipschitz function with the same Lipschitz constant $K$ than $R$, for each element $s \in M_0$
we have
$$
|R^M(s^*) - R(s)| \le   K d_\epsilon (s^*,s') .
$$
Fix now $s \in M_0.$ Then by hypothesis there is a functional $a_s \in \mathcal A_{M_0,R}$ such that
$$
|R^M(s^*)- s^* \cdot a_s| =
|R^M(s^*)- s \cdot a_s + (s-s^*) \cdot a_s|
$$
$$
\le
|R^M(s^*)- s \cdot a_s| + |(s-s^*) \cdot a_s| \le K d_\epsilon(s^*,s) + |(s-s^*) \cdot a_s|.
$$
Therefore,
$$
|R^M(s^*)- s^* \cdot a_s |  \le  K \big( \Theta(s^*,s) + \epsilon \|s- s^* \|_2 \big) +
 100 \, \|s-s^*\|_\infty.
$$
Since this happens for all the elements $s \in M_0$, we have that the inequality holds for the infimum. Finally, note that  the set $M_0$  is compact. Indeed, by Lemma
 \ref{lema} $d$ and $E$ are equivalent metrics on $M$.
 we have that the infimum is attained, and so we get the result by taking $a_{s^*}= a_{s_0}$ for the state $s_0$ that attains the minimum.
}

\vspace{0.3cm}

Using this result with some restrictions on  the set $M_0$ and the relation with the particular
elements $s^*,$ we obtain useful bounds for the formulas that approximate $R^M.$ We write one of
them in the next corollary. Essentially, it reflects what happens with the extension of the reward function $R$ for a state $s^*$ that represents the same market trend as another state belonging to $M_0,$
but with different norm.

\vspace{0.3cm}

\begin{corollary}
Let $M_0 \subset M$ be a compact subset of $(\mathbb R^n \setminus \{0\},\|\cdot\|_2)$. Consider a function $R:M_0 \to \mathbb R$ satisfying the requirements in Proposition \ref{propext}.

Suppose that an element $s^* \in M$  belongs to $\{ \lambda >0: \lambda M_0\}.$
Then
there is a functional $a_{s^*} \in \mathcal A_{M_0, R}$ such that
$$
|R^M(s^*)- s^* \cdot a_{s^*} | \le
\min_{ 0 < \lambda} \, \Big\{ \frac{|\lambda-1|}{\lambda} \Big(  100 \, \|s^*\|_\infty + \epsilon K \, \|s^*\|_2 \Big): \, \, \frac{ s^*}{\lambda } \in M_0   \Big\}.
$$
\end{corollary}
\textit{Proof. 
By assumption, $s^*= \lambda s$ for a given $0< \lambda$ and $s \in M_0.$ For such an $s$
we have that $\Theta(s,s^*)=0.$ The rest of the right hand term in the inequality in
Proposition \ref{propext} can be rewritten as
$$
100 \, \|s-\lambda s \|_\infty +  \epsilon K E(s,\lambda s) \big)
=
|1-\lambda| \, \|s\|_\infty )+ |1-\lambda| \, \epsilon K \|s\|_2,
$$
for $s^*=\lambda s.$ This can be rewritten as
$$
\frac{|\lambda-1|}{\lambda} \Big(  100 \, \|s^*\|_\infty + \epsilon K \, \|s^*\|_2 \Big).
$$
This gives the result.
}

\vspace{0.3cm}

Depending on the geometry of the set $M_0$ and its relation with the chosen state $s^* \notin M_0,$ we can also obtain a lower bound for the approximation formula of $R^M$
using actions $a \in \mathcal A.$

\begin{proposition} \label{lowerb}
Let $M_0 \subset M$ be a compact subset of  $(\mathbb R^n \setminus \{0\},\|\cdot\|_2)$, and $\epsilon >0$.
Consider a function $R:M_0 \to \mathbb R.$
Let $s^* \in M \setminus M_0$ and $a \in \mathcal A$ such that
$$
s^* \cdot a \ge R(s) \quad  \textit{for all} \quad s \in M_0.
$$
Then for
$\Theta(s^*,M_0) = \inf_{s \in M_0} \,\Theta(s^*,s)$ and $E(s^*,M_0) = \inf_{s \in M_0} \,\|s^*-s\|_2,$ we have that
$$
\big| s^* \cdot a - R^M(s^*) \big| \ge K \big( \Theta(s^*,M_0) + \epsilon E(s^*,M_0) \big).
$$
\end{proposition}
\textit{Proof.
Take $s^*$ and $a \in \mathcal A$ as in the statement of the result. Then, using again compactness of $M_0$ we get an element
$s_0 \in M_0$ such that $R^M(s^*)=R(s_0)-K \, d_\epsilon(s_0,s^*).$ 
So we have that
$$
\big| s^* \cdot a - R^M(s^*) \big| = \big| s^* \cdot a - R(s_{0}) + K \, d_\epsilon(s_0,s^*) \big|
$$
$$
= \big( s^* \cdot a - R(s_{0}) \big) + K \, d_\epsilon(s_0,s^*) \ge
 K \big( \Theta(s^*,M_0) + \epsilon E(s^*,M_0) \big),
$$
and the lower bound is proved.
}

\vspace{0.3cm}

In particular cases, this bound can be used for getting clear negative results on the possibility of approximating the extended reward function $R^M$
by means of actions. We show two of them in the following result, which proof follows the same easy arguments than in Proposition \ref{lowerb}. Note that the distances point-to-set $\Theta(s^*,M_0)$ and  $E(s^*,M_0)$ defined above make sense if $M_0$ is changed by any other set.  

\begin{remark} \label{speciallowerb}
Let $M_0 \subset M$ be a compact subset of  $(\mathbb R^n \setminus \{0\},\|\cdot\|_2)$, and $\epsilon >0$.
Consider a function $R:M_0 \to \mathbb R,$ and let $s^* \in M \setminus M_0$ and $a \in \mathcal A.$

\begin{itemize}

\item[(i)]
If
$s^* \cdot a \ge 100 \, \|s\|_2$ for all $s \in M_0,$ $\sup_{s \in M_0} \|s\|_2=B$
and $s^* \in \cup_{\lambda >0} \lambda M_0,$ then
$$
\big| s^* \cdot a - R^M(s^*) \big| \ge K \epsilon \big( \|s^*\|_2 - B \big).
$$

\item[(ii)] If $M_0 \subset C,$ where $C$ is a closed  convex cone (with vertex  $0$) that does not contain $s^*$, and let 
$s^* \cdot a \ge R(s)$ for all  $s \in M_0.$ Then
$$
\big| s^* \cdot a - R^M(s^*) \big| \ge K \, \Theta(s^*,C).
$$
\end{itemize}

\end{remark}

\begin{remark}
As we have demonstrated, the mathematical model imposes the restriction that valid states are always different from $0.$  That is, there are no states that represent that the system has not changed, or that there is no trend. Therefore, these states have to be eliminated if they appear in the experience.
\end{remark}

\vspace{0.3cm}

\section{Designing models for financial markets using Lipschitz extension-based reinforcement learning}
\label{S3bis}

We will show in this section how to proceed to use our mathematical framework. Essentially, as we said, we use the previous steps of a dynamic process to compute an extension of a reward function---a Lipschitz function---, which allows us to calculate which is the best action of a given subset given to execute in the next step. We will present two complete models, with the aim of showing the scheme to follow to build a model following our technique. We will check the results in Section \ref{S4}, analyzing the total reward obtained by using each algorithm, comparing the result with the optimal reward computed \textit{a posteriori}, and with other methods.

\subsection{ An expert system for day-to-day investment in a currency market product} \label{subS3.1}

We explain in what follows a canonical example of application of our technique. Suppose that we are interested in investing in the change of a given electronic currency in US dollars, for example, Ethereum. Data were downloaded from the web page https://finance.yahoo.com/cryptocurrencies, 31 August 2019.
We want to use the support of an algorithm every day in the starting moment of the session, and we want to decide how much money we could invest, whether we should buy or sell the currency we have, and whether we should reserve part of the money we have for intraday investment (buy/sell several times in the same day, following the instant sign of the market).

\vspace{0.3cm}
 
(A)
We will consider the dynamic in a full year. We fix that the relevant information in a given day is given in a vector of 3 coordinates, containing as first coordinate the balance of the previous day (that is to say, the difference between the value of closing and the value of opening). The second coordinate is the initial value at today's opening divided by $100,$ and the third is the total amount of investments recorded in the market the previous day divided by $10^8.$ The weights are included to avoid large differences in the absolute values of the investments. 

However, for computing the value of the reward function after each day---that is, the true value of the investment done---, we need  to define other set $D$, given by the balance at the end of the day, the maximum variation during the day, and the relative global volume of investment. We call to the vectors of $D,$ once they are known, the state-values associated to the states. The set of action have to be interpreted as scalar functions over these states acting by duality. The actions represent  investing strategies  as  3-coordinates vectors that are bounded in norm: the sign of the coordinates indicates if the action must be positive or negative; for example, if the fist coordinate is positive, then the decision maker is advised to buy. The size of each coordinate represents how much effort is recommended to put into each of the three action items. As explained in the previous section, the reward associated to each proposed action is the scalar product of the state and the action. For already known states ---that is, for states that have already passed---, this is given by the scalar product of the state-value vectors of $D$ associated to the given state and the action. The optimal reward for states that have already passed, that will be used for comparing the results of the model, is computed by maximazing the state-value vector with respect to the set of actions chosen, that gives the $2$-norm of the state-value vector if the actions considered are all the vectors in the unit ball $B_{\ell^2}.$

\vspace{0.3cm}

(B) 
The learning procedure is designed to be a step by step extension of the reward function. We consider a fixed set of actions that is defined as a randomly chosen set $\mathcal A$ of $30$ norm one vectors. 
The first day (k=1) we have no information yet, so we start the second one. For this day, the metric space $M_0$ has only one element $s_1,$ what is enough to compute the Lipschitz extension for the next day.  The reward function associated to $s_1$ can also be computed: it is the norm of  $d_1,$ the  element of $D$ corresponding to the first day. The optimal action $a_1$ that allows to compute the reward as $d_1 \cdot a_1$ is the norm one 3-coordinates vector that gives the maximum, that is, the norm of $s_1.$ The procedure follows in the step $k$ by adding the every day experience from the first day and the day $k-1.$

\vspace{0.3cm}

(C) At the step $k,$ the estimate of the reward function $R^M$ is constructed by applying the McShane extension formula to the reward function $R$ given by the rule $(s_i,a_i) \mapsto d_i \cdot a_i,$ $i=1,...,k-1.$
That is, after computing the corresponding Lipschitz constant $K,$ the formula 
$$
R^M(s_k):= \sup_{s\in M_0}\{R(s)-K\,d(s_k,s)\}, 
$$
provides the value of the desired extension $R^M: M \to \mathbb R$ to the action $s_k.$  Instead of the McShane formula other extension rules ---such as Whitney formula--- can be used (see for example, \cite{rao} and the references therein). Of course, the results will depend on this choice; some comments on why we have chosen this option are presented in Remark \ref{remrew}, but the use of other options is open. Checking which is the best extension formula exceeds the scope of this paper, which is only intended to introduce the technique.

At this step, the set $M_0$ is defined by the product of the states $s_1,...,s_{k-1}$ with the product distance $d_s + d_a$, where both metrics are defined  
as in (\ref{epdistance}) for $\epsilon =1/10,$ in the space of states of $k-1$ elements and the  space of actions of $30$ elements. Note that the only pair $(s,a)$ which has a clearly positive reward is given by the state $s_i$ together with the action $a_{s_i}$ that allows to compute the best reward associated to $s_i,$ that is $d_i \cdot a_{s_i}.$ Thus, the rest of the pairs $(a,s)$ together with their rewards are artificially created, and define what we call \textit{dreams}, concerning in this case the creation of new pairs (state,action) together with the correponding rewards, although the ``state" part
comes from real data. The new training data that will be called dreams in the next section are constructed by directly creating new states. 

The state $s_k$ is then considered. All the possible forecasts of the rewards for the pairs $(s_k,a),$ $a \in \mathcal A,$ calculated by using $R^M,$ are computed. The maximun of this set is computed, and the action $a_{k_0}$ for which this maximum is attained is chosen to be the solution at this step: at the begining of the day $k,$ the decision-maker has to choose the action $a_{k_0}.$

In order to study the effectiveness of the process, we will consider the cumulative reward foreseen by our technique, in order to compare it with the cumulative optimal reward. This will be done in  Section \ref{S4}.

\vspace{0.2cm}

\subsection{Investments distributed in several products of a stock market}    \label{subS3.2}

\vspace{0.3cm}

In this case, we will work with the following metric space structure as a model for the dynamical system defined by a financial market with $n$ products. We will assume that there are $m$ times in which there are share purchase/sale events. The model tries to solve the problem of distributing the daily investment in $4$ different products, indendently of the total amount to invest. The previous $k-1$ steps of the time series is used to define the metric space in which the reward function acts, and to compute an expected value of the function for the step $k.$ 

\vspace{0.3cm}

(A)
Take a  subset  $M_0$ of vectors of  $M=\mathbb R^n$ representing the states of the market. Each of the vectors in $M_0$ describes a
state of the market in the following way: each coordinate gives the value of the \textit{increment} of the corresponding product at this moment. In fact, we will write at each coordinate $i$ the difference of the value at the moment $i \in \{1,...,m\}$ and the value at $i-1.$
This means, in particular, that the original values of  the products is not relevant for defining the states, just the variations.

Again, we will fix the value $\epsilon=1/10$ for the definition of the metric in the next sections. That is, we will use
$$
d(s_i,s_j)=d_{1/10}(s_i,s_j)= \Theta(s_i,s_j) + \frac{1}{10} E(s_j,s_i), \quad s_i, s_j \in M_0.
$$

By choosing  the value $\epsilon=1/10$ we try to obtain a balance among the ``norm part” and the ``angular part". We estimate that the values of the differences among vectors with respect to the norm is around $10,$ while the maximum value of the angular part is $1.$ A more systematic way of calculating this parameter would be necessary in a more advanced version of our method.

 \vspace{0.3cm}

(B)
We are interested in measuring the success of a concrete action in the market,  that  is, the success
of a  share purchase/sale event that a decision-maker has executed on the market. So we have to define what an action is in the model. Formally, we have already defined them as elements of the dual of
 $\mathbb R^n.$ As we said, at each step the state of the system is defined by an
 $n$-coordinate vector; each coordinate represents the increase/decrease of the value of each product with respect to the previous
step. An \textit{action} is a suitable share purchase/sale event that the decision maker could execute, represented as follows: it is supposed that he has 100 monetary units to invest at every step, so an action
is a vector of $n$-coordinates ($n+1$  if we want to consider leaving some of the money out of the buying process). In Section \ref{S4} we will call ``bets" to the actions to reinforce their meaning in the model.
  Mathematically, they are  positive elements of the algebraic dual of $\mathbb R^n$ having $\ell^1$-norm equal to $1$. Let us write $\mathcal A$ for the set of all the actions.

The natural reward function to be defined in the model must be related to the  evaluation of the success of an action when it is applied to a certain state of the system. Therefore, it must be defined as a functional acting in $\mathcal A$ once a given state of the system has been fixed, and so it is a two-vector-variables function $R_0$  acting in $M_0 \times \mathcal A.$

However, the reward function must evaluate   states of the market ---an element of $M_0$---, taking into account how the decision maker acts in it and the success of his actions. Therefore, we will finally consider  a reward function $R$ acting in $M_0$,
but we will use all the information we have about the system to estimate it. That is, we will use  the function $R_0$ for defining the function $R$.
We will see that, finally, for each state $s \in M_0$ there is an action $a \in \mathcal A$ such that $R(s)=s \cdot a,$ or a mean of such values.

 \vspace{0.3cm}

(C)
After this,
 we are interested in  extending  the reward function $R$ to the whole linear space $M$ preserving the Lipschitz constant. In order to assure that this constant is a  (positive) real number,
 it is enough to take into account that the set $M_0$ is defined by a finite set of vectors. In the model, $R$ is supposed to measure ``how successful" is a given state.
We will use  the McShane formula for the extension. The extension $R^M$ is supposed to extrapolate the same concept ---success of a given state---, preserving the metric relations of $M_0$ and $M$.
Since it appears explicitly in the formula, we have to compute the Lipschitz constant $K$ for the reward function $R$ in order to get the extension
$R^M,$ for which the same $K$ works. The way we have defined the metric in the space allows to obtain a theoretical bound for this extension, as stated in Proposition \ref{propext}. However, note that in general we cannot expect that $R^M(s)$ can
be represented as an action belonging to the positive part of $100 \, \times \, B_{(M,\|\cdot\|_{\ell^1})}.$ This was  shown in the Example of Section \ref{S3}; the general behavior of such kind of representation formula was discussed also there, as a consequence of Proposition \ref{lowerb} and its corollaries.

\vspace{0.3cm}

(D)
Finally, we will use $R^M$ for simulating the reward of new time sequences of states in order to perform our reinforcement learning algorithm.  In order to do this, we generate randomly
new states for increasing the set $M_0$. We create in this way a new seminal set  $M_1$ bigger than $M_0,$ in which we are mixing ``known situations" ($s \in M_0$)  and new ones,  that we call ``dreams" ($s^* \in M_1 \setminus M_0.$) The rate of elections of known cases and dreams that we have chosen is $\beta = 50 \%.$


\vspace{0.3cm}

\section{Training and dreaming: Lipschitz approximations to  real market reward functions
for  the design of reinforcement learning algorithms} \label{S4}

\vspace{0.3cm}

We will present in this section an analysis of the application of the method explained before, comparing the results provided by our algorithms and other standard techniques and the optimal strategy computed \textit{a posteriori.}

\subsection{Efficiency analysis of  Lipschitz extensions in a currency market}

Following the procedure explained in  \ref{subS3.1}, we fix first the relevant sets that have to be used for testing our method. 
The description of each state, at the day $k=1,...,365,$ is given by the 3-coordinates vector $s_k=(s_{k,1},s_{k,2},s_{k,3}),$
\begin{align*}
	 & s_{k,1}= dataOpen(k-1)-dataClose(k-1), \\
	&s_{k,2} = dataOpen(k) \cdot 10^{-2},   \qquad \textrm{and}\\
	&s_{k,3} = dataVolume(k-1) \cdot 10^{-8}.
\end{align*}
where \textit{dataOpen} and \textit{dataClose} are the values at the beginning and at the end of the day on the exchange market,  and \textit{dataVolumen} is the volume of business in the currrency market,
both of them of the day before of the one considered, that is $k.$

The 3-coordinates vector $r=(r_{k,1},r_{k,2},r_{k,3})$ gives the description of the real results obtained this day, that are known after each day has passed. 
\begin{align*}
& r_{k,1}= dataOpen(k)-dataClose(k), \\
& r_{k,2} = ((dataHigh(k)- \max(dataOpen(k),dataClose(k))) \\
& \,\,\,\,\, \,\,\,\,\,  -(dataLow(k)-\min(dataOpen(k),dataClose(k)))) \cdot 10^{-2},   \quad \textrm{and}  \\
& r_{k,3} = (dataVolume(k)-meanVolume) \cdot 10^{-8},
\end{align*}
where \textit{dataHigh} and \textit{dataLow} the highest and the lowest exchanges in the day,  and \textit{meanVolume} the average value through all the year, or of the previous year. We use this vector to compare the results of our technique and the neural network constructed for the aim of comparison. The vector $r$ provides also, just by maximizing the reward, what is the best action to apply, that of course can be known only once the day has passed. This allows to find the real reward, that would be obtainied in the ideal situation that every day the decision maker invests using the best action.

The set of actions is obtained by choosing randomly thirty elements of the set $S_{(\mathbb R^2,\| \cdot\|_2)} \times \{-1,+1\}$, that is, 3-coordinates vectors where the two first components belonging to the sphere of the 2-dimensional  Euclidean space, and the third coordinate being $1$ or $-1.$

To check the method, we compute the cumulative reward that results when we apply our algorithm to the daily investment in a currency exchange market. On the other hand, we also train three neural networks with a complete set of  experiences in the same market to solve the same problem. Finally, and using the information at the end of the day in all the steps of the time series, we compute the cumulative reward using the optimal investment option, that was calculated as the sum of the  norms of the corresponding elements of the set $D$ as explained in \ref{subS3.1}. The result is presented in the following  Figures \ref{RplotRewAc}, \ref{RPlotPint} and \ref{Rplot50_100_150}.

 \begin{figure}[h]
\caption{ Ideal exchange market benefits: cumulative reward by taking the best possible sequence of investments
.}
\vspace{0.5cm}
\centering
\includegraphics[width=0.8\textwidth, height=0.25\textheight]{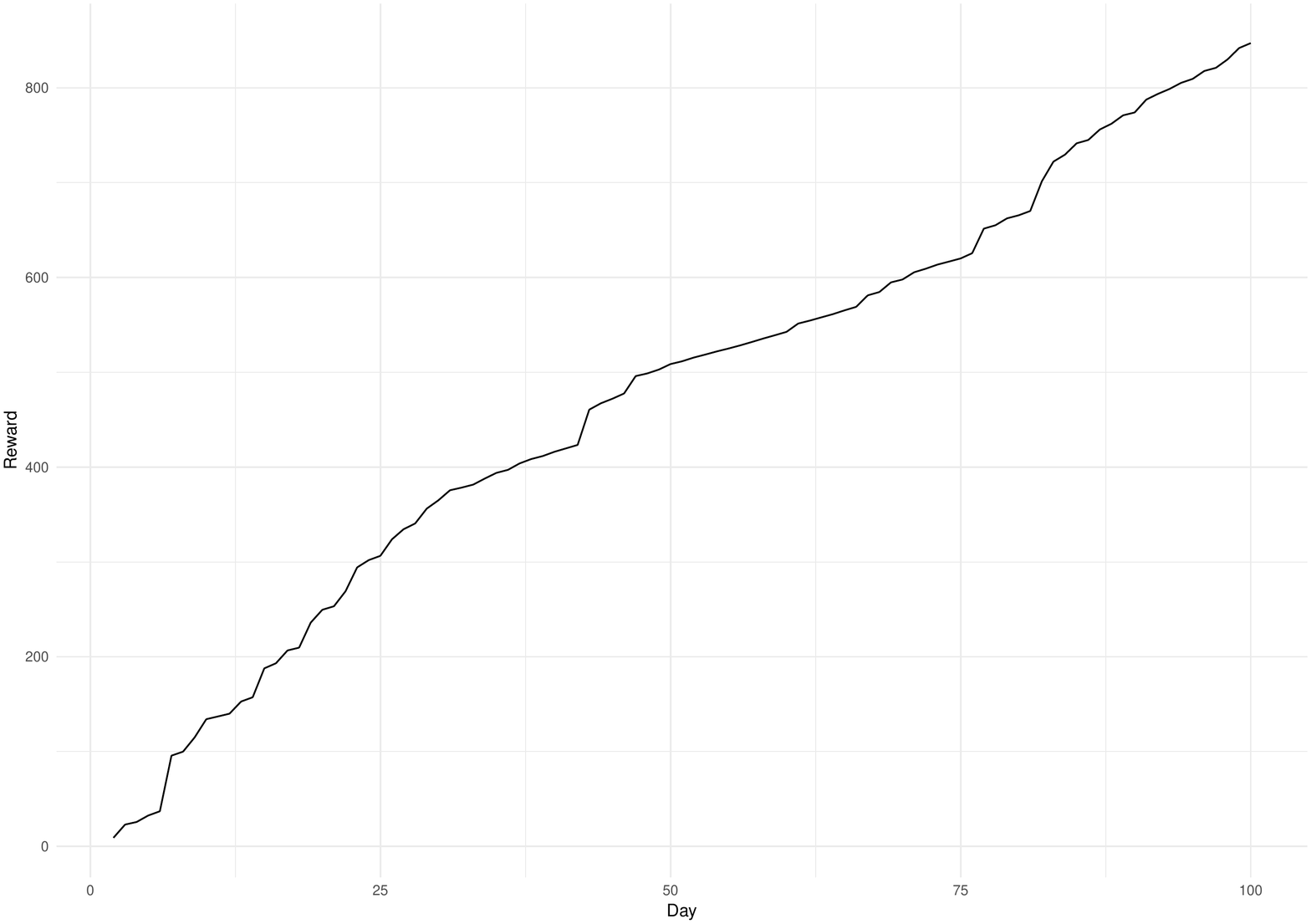} \label{RplotRewAc}
\end{figure}

 \begin{figure}[h]
\caption{ Benefits in the exchange market with our method: cumulative reward.}
\vspace{0.5cm}
\centering
\includegraphics[width=0.8\textwidth, height=0.25\textheight]{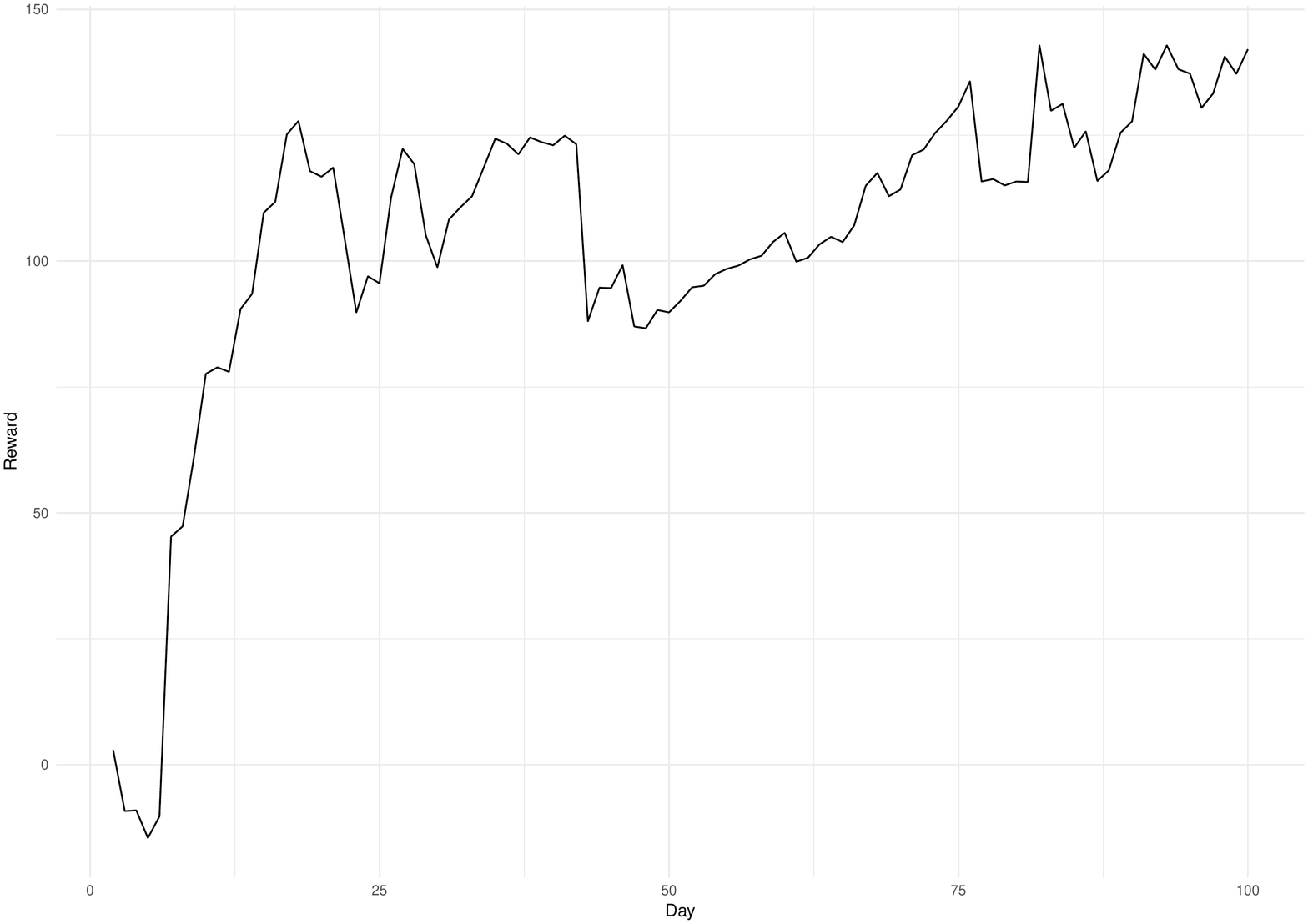} \label{RPlotPint}
\end{figure}

\begin{figure}[h]
\caption{ Neural networks: cumulative reward. (Black=Dataset Size 50, Red=Dataset Size 100, Blue=Dataset Size 150)}
\vspace{0.5cm}
\centering
\includegraphics[width=0.9\textwidth, height=0.3\textheight]{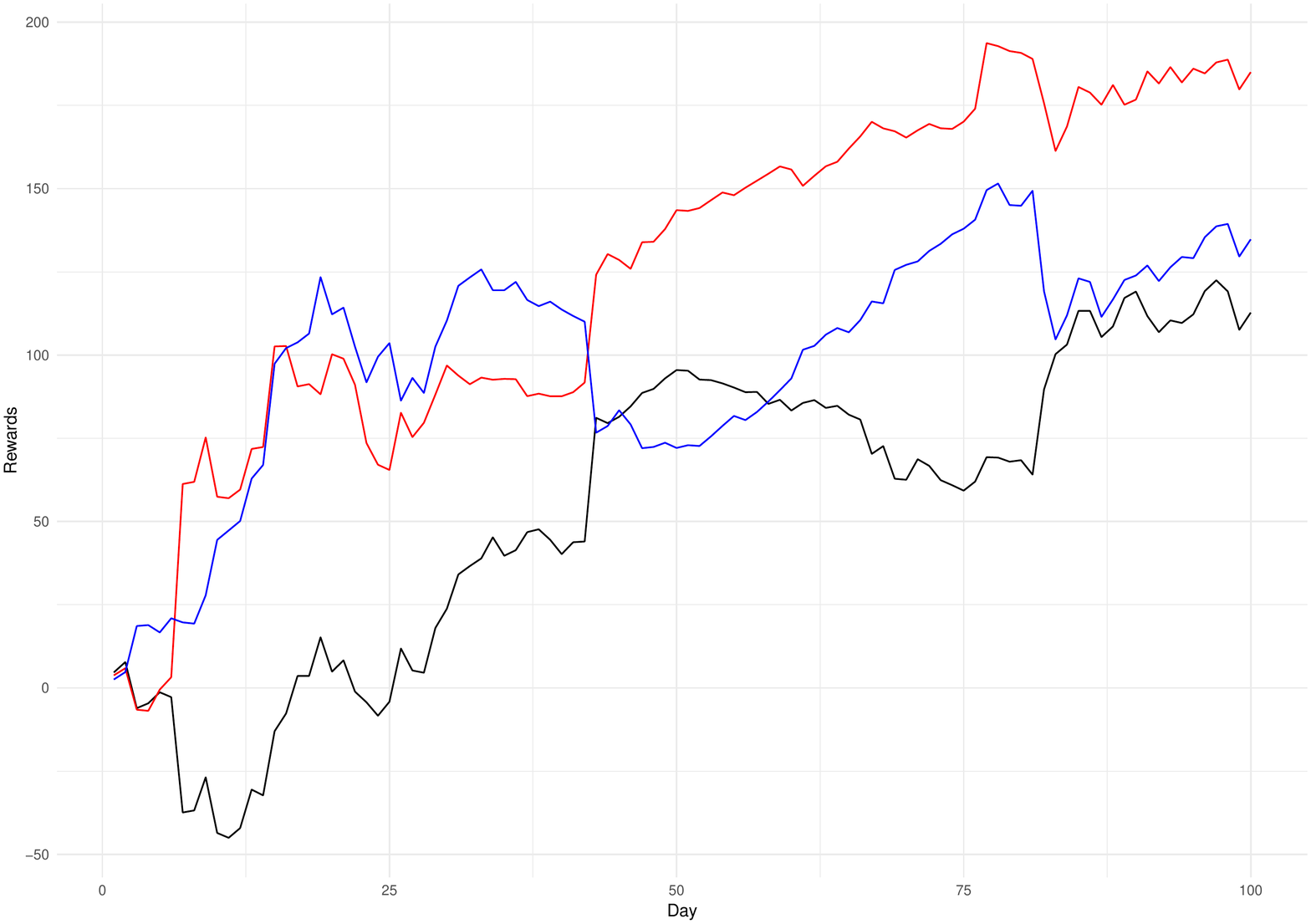} \label{Rplot50_100_150}
\end{figure}

The procedure for constructing an alternative neural network-based method for making the prediction uses a mixed scheme that introduces some reinforcement learning ideas.  We consider the same elements as those used in the Lipschitz extension procedure, i.e. vectors representing states as explained in \ref{subS3.1} defined by the variables \textit{dataOpen}, \textit{dataClose} and \textit{dataVolume}. With the same reward function considered above, we calculate the optimal action to be executed by the decision maker to obtain the best benefit as in the previous case, by duality with the vector of real results of each day. The neural networks, one for each coordinate, are trained using the vectors that describe the states to obtain the corresponding coordinate of the optimal action. 
We define training sets from some past experience, associating to each state of this experience (a set of states from the previous year) its optimal action. 

Three cases are considered, with three neural networks in each. The architecture of the neural networks is the same in all cases: one layer with a variable number of neurons.   In the first case, a data set corresponding to 50 days of previous experience and 40 neurons ---in the 3 networks trained for the 3 coordinates---, are considered. The second case uses a data set of 100 days, with 70 neurons, and the third 150 days with 100 neurons. At each step, the new training datasets contain the previous ones. The results can be seen in Figure \ref{Rplot50_100_150}. 

%
%

\vspace{0.4cm}


As can be seen, the algorithm we propose gives easily interpretable results (Figure \ref{RPlotPint}). Some of the values ---one in ten---  are presented in Table \ref{tabla1}. First of all, it can be seen that the comparison with the optimal investment shows a reasonable level of success.  The first steps give a negative cumulative reward, due to the lack of experience, but soon begins to improve, giving in the first 20 steps a return of about 50\% of the optimal reward. The increase is slow after this moment; however, it can be seen that the cumulative reward is sometimes decreasing, but the global behaviour is increasing. 

Neural networks also give a positive cumulative reward, although the results are worse in two of the three cases considered. It should be noted that neural networks are trained from the first moment with data corresponding to actual experience of at least 50 days, while our method uses experience data of $k-1$ days for the day $k.$ The reader can compare the results with the help of  Table \ref{tabla1}, in which predicted values for trained neural networks with data sets ---containing information on 50, 100 and 150 days--- can be seen in the first 3 columns, respectively, as well as the results for our method (fourth column) and the calculated optimal reward \textit{a posteriori} (fifth column).
Also, keep in mind that the McShane formula penalizes rewards calculated from data that are very far (with respect to the distance we have defined) from the state being considered: it is a maximum with a negative term that is defined exactly by this distance. In the case of neural networks, this coherent behavior cannot be expected in general, since one cannot have any control over the free parameters of the fitting method.

\vspace{0.5cm}

\subsection{Creating dreams for the reinforcement learning procedure in parallel investments}

Let us continue with the presentation of the procedure by further specifying the example explained in \ref{subS3.2}.
Suppose that we are analyzing  a fixed market with four similar products. In fact,  the dynamics of their prices 
are equivalent, as the reader can see from the figures below.
We have the complete  sequence  of their values  each minute  from $1$ to $800.$  As we said in Section \ref{subS3.2} and for the aim of simplicity, we assume that at the beginning of the process the values of all the products equal  $0$.
The  set $M_0$ of known states for which the reward function is known is considered to be the first half
of the states that have been registered. Let us see how we fix the mathematical representation of the problem.


\vspace{0.2cm}

(1)
A state of the system is given by a four-coordinate vector $s$: as we explained in Section \ref{S3}, each minute the vector gives the cumulative increase or decrease of the values of each product. Since we want to define the reward function using the scalar product with a vector that represents an action, and we want to include the possibility of not investing, we expand the vectors $s$ by adding a fifth null coordinate. We preserve the same symbol $s$ for the extended vector.

 \begin{figure}[h]
\caption{ Real market experience: set of states for training the model. The cumulative values for all the products of the market are represented.}
\vspace{0.5cm}
\centering
\includegraphics[width=0.9\textwidth, height=0.3\textheight]{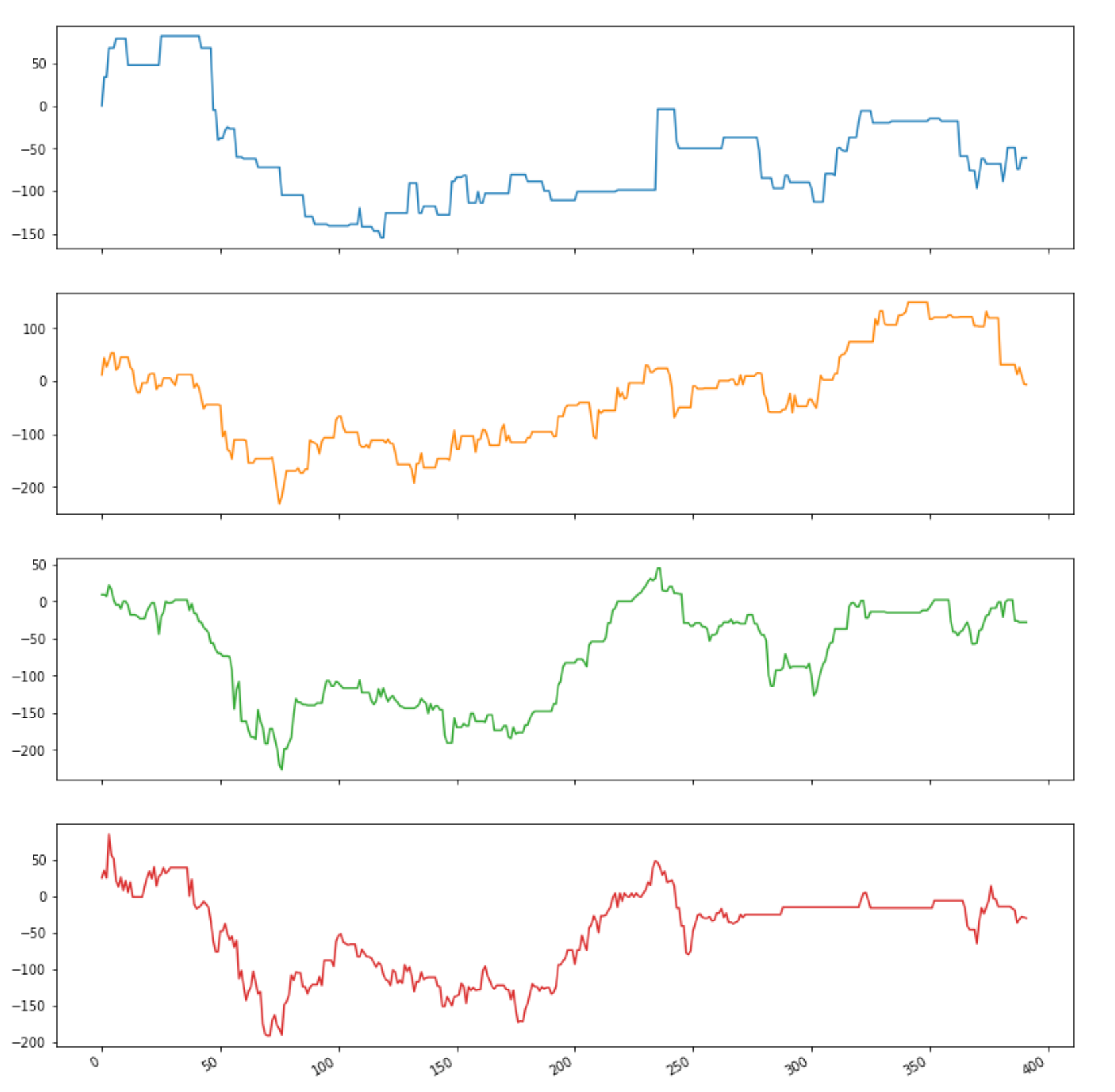} \label{training}
\end{figure}

We consider series of ``bets" applied at each minute. They correspond to series of what we called ``actions" in \ref{subS3.2},  that in this particular case  are described as the $\%$ of the money  that the decision maker wants to apply in each market  this minute (including not investing a certain part). The decision maker is investing $100$ monetary units at each step. 

\vspace{0.2cm}

(2)
Fix now a (five-coordinate) state of the system $s$.
The reward function $R: M_0 \times \mathcal A \to \mathbb R$ is then defined as a two-(vector)-variable function given by  the scalar product of the state $s$ and  the action $a$,
$R_0(a,s)=  \, a \, \cdot \, s.$

At this point we introduce our first arguments regarding reinforcement learning. The main idea is to use  the information that is known for similar situations in order to compute a reward function $R:M_0 \to \mathbb R$, depending only on the state. Note that this is different that was done in the other case introduced in Section \ref{subS3.1}, since in this case the reward function and its extension was defined considering pairs (state,action). This is relevant, since we are going to evaluate the state of the system using this reward function.
In order to define it, we use the following procedure. For a state of the system $s,$
 we define
$$
R(s):= \, mean \, \{ \, R(a,s) \, : a \in  A \cup B \},
$$
where the average is calculated on two sets of $A$ and $B$ built as explained in what follows, whose sizes are in a ratio of $90 \%$ and $10\%,$ respectively. The first set $A$ ---$90  \%$--- is defined by using  actions/bets $a$ that have been already checked and
 have obtained good enough values of the reward functions when acting in states $s'$ that are similar to $s$.
This is done by choosing the bets that give the highest values of the reward function when they act on these states $s'.$
The similarity relation is given by proximity with respect to the distance $d$, that is $d(s,s') < \varepsilon$
for a given $\varepsilon >0$ (for example, $\varepsilon =0.5$).
The second set $B$ ---$10 \%$--- is randomly obtained.

%

\begin{figure}[h]
\caption{Sequence of (randomly chosen) actions that optimize the bets when applied to the set of real states $M_0.$ Note that for each fixed time, the five values equals $100 \%.$}
\vspace{0.3cm}
\centering
\includegraphics[width=1.0\textwidth, height=0.3\textheight]{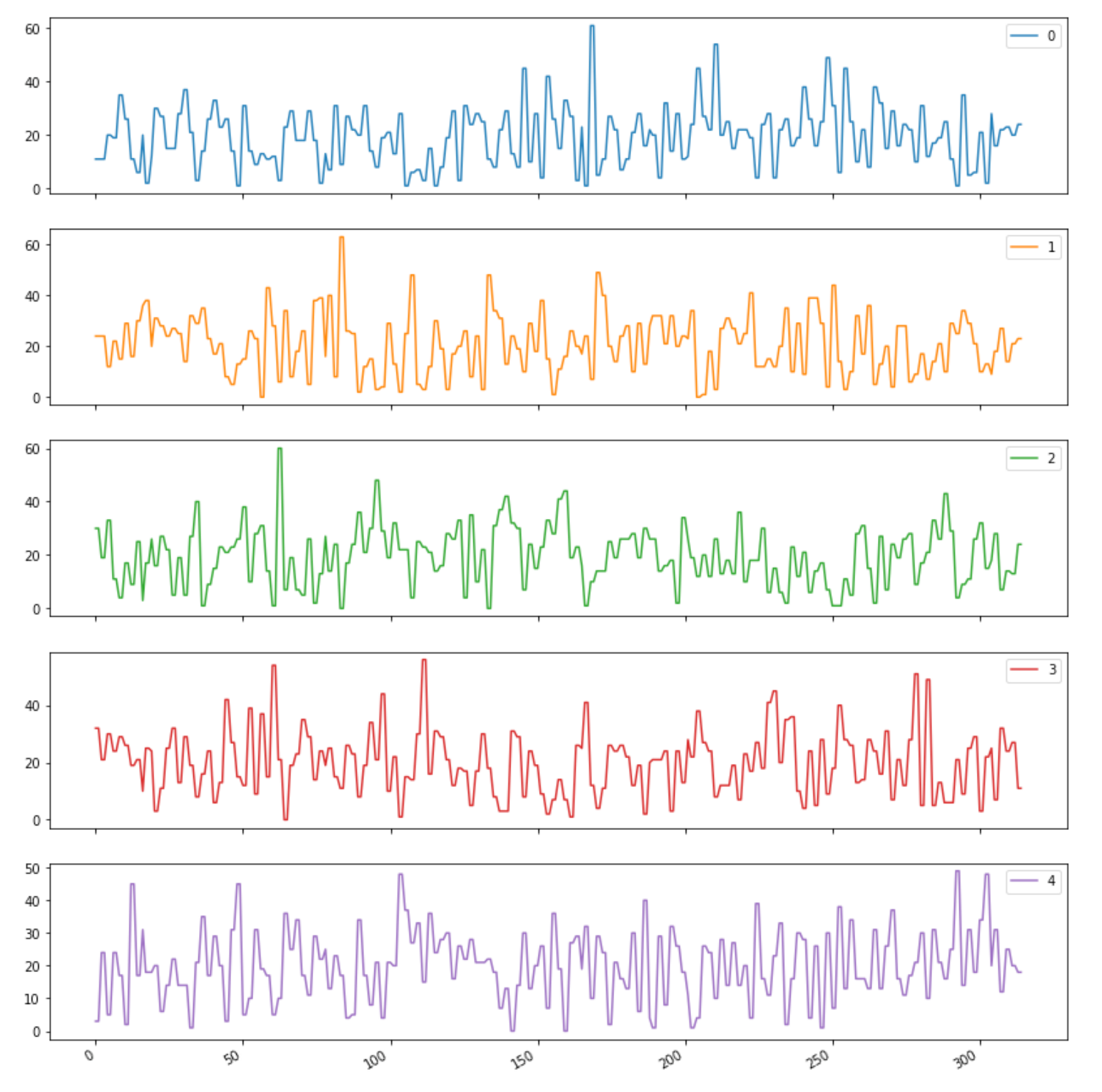}  \label{allocations}
\end{figure}

%


This method is used for computing the reward function $R$ for the elements of $M_0$. As we said 
in  Section \ref{subS3.2}, for states which do not belong to $M_0$---i.e. for the remaining $50 \%$--- we will use the McShane formula for obtaining the extended function $R^M.$

%

\hspace{-0.5cm}
\begin{figure}[h]
\caption{Simulation with real data obtained from the experience.}
\vspace{0.3cm}
\centering
\includegraphics[width=1.0\textwidth, height=0.25\textheight]{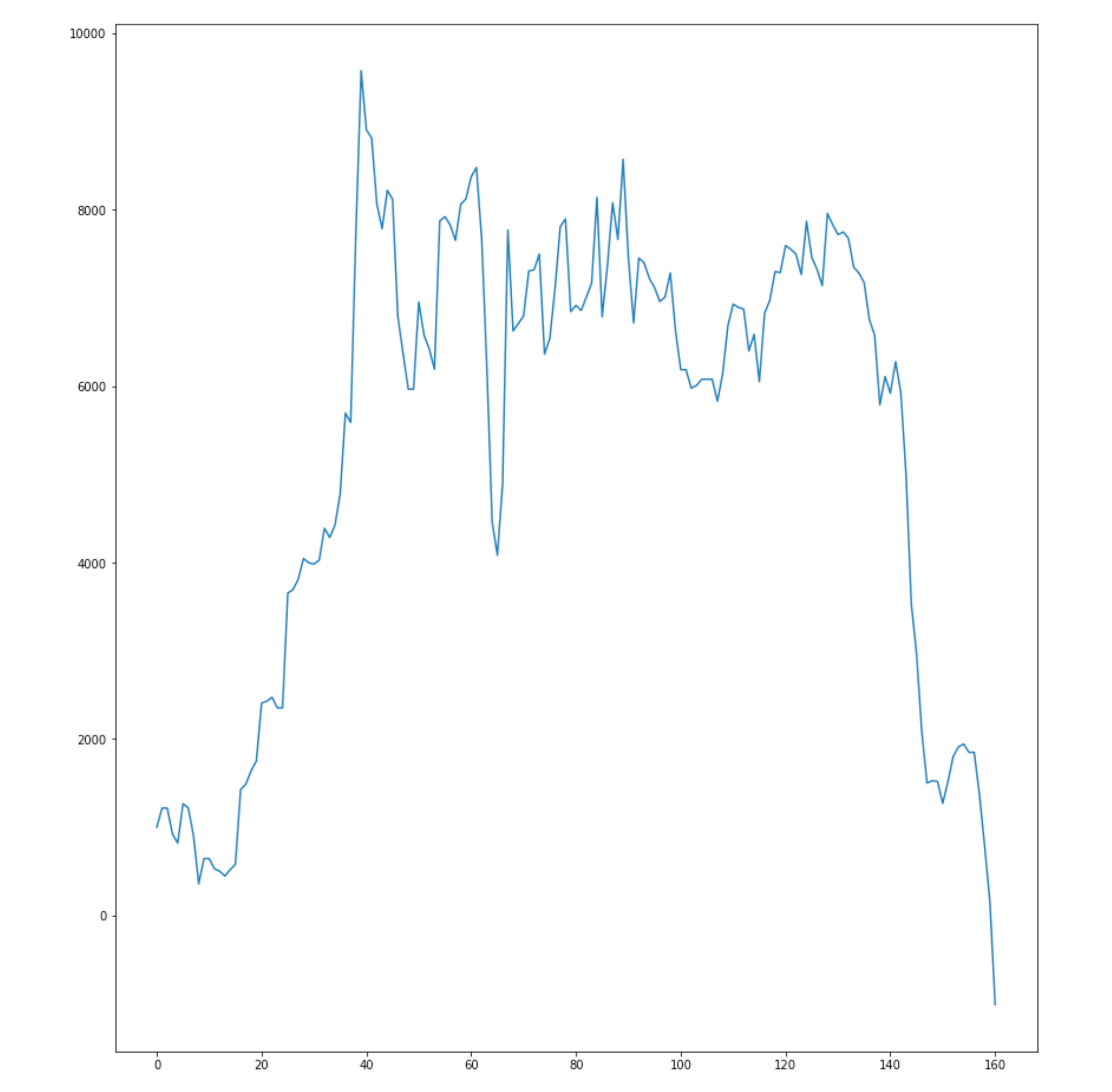}  \label{CUMSUMnodream}
\end{figure}

\hspace{-0.5cm}
 \begin{figure}[h]
\caption{Simulation with $50 \%$ of real data +$50 \%$ of dream.}   \label{CUMSUMwithdream}
\vspace{0.3cm}
\centering
\includegraphics[width=0.9\textwidth, height=0.25\textheight]{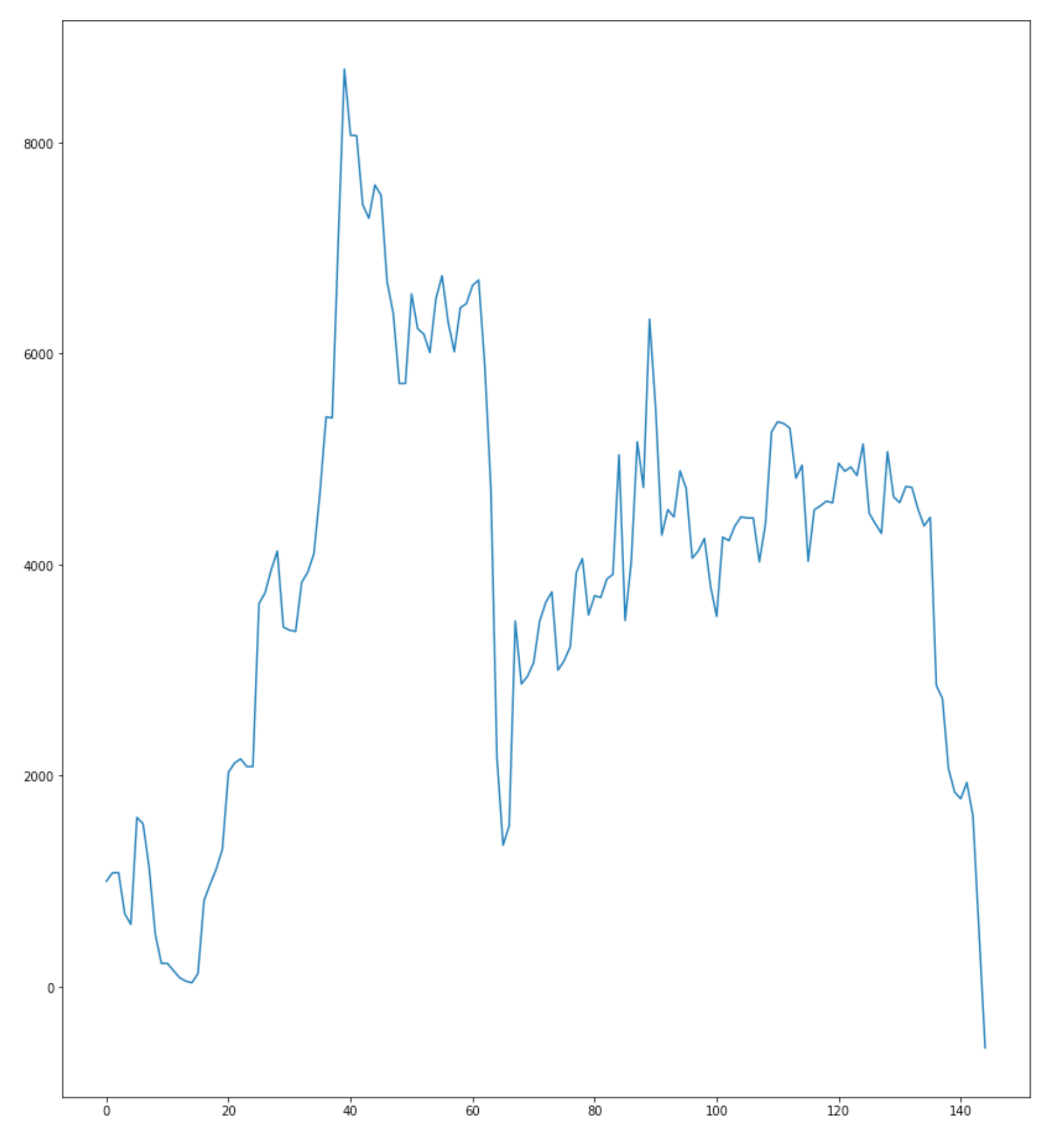}
\end{figure}

\vspace{0.4cm}
(3)
Taking into account the procedure for obtaining the reward function $R$, given a state $s \in M_0$ we can find an action/bet $a_s \in \mathcal A$ such that $R(s)=R_0(a_s,s)$
and is as good (high) as possible, as was done in the case explained in \ref{subS3.1}. Of course, $a_s$ is not unique, but one of the possible solutions can be fixed using a random procedure.

%

A similar definition can be done for suitable states that do not belong to $M_0$. We call dreams to such states. In this case, the reward function that should be considered is $R^M,$ since this function plays the role of $R$
for states that have not been found in the experience in the market. However, note that we cannot say that, if $s^* \in M \setminus M_0,$ there is a positive functional ---an action--- $a_{s^*}$ in the unit ball of $\ell^1$ such that $R^M(s^*)= s^* \cdot a_{s^*},$ as happens for $s \in M_0$ and $R.$ This problem is solved just by taking a suitable  ``norm $100$" functional $a_{s^*}$ such that $R^M(s^*) - s^* \cdot a_{s^*}$ attains its minimum value. We have already proved that in general, $R^M(s^*)$ cannot be attained by a value as $a_{s^*} \cdot s^*.$
However, Proposition \ref{propext} and the results in Section \ref{S3}   give precise bounds for this difference.

The set of all ---randomly chosen but optimal--- bets as $a_s$ and $a_{s^*}$ represents how the decision maker should act when he faces the problem of investing in the market.
Figure \ref{allocations} shows a representation of a suitable set of optimal bets for the states represented in Figure \ref{training}. 


As we have shown, the main tool of our technique is the computation of the McShane extension of the reward function. In order to clarify this computation based on the McShane-Whitney extension theorem, we provide  an scheme of the algorithm (Algorithm \ref{alg:algoritmRM}).

\vspace{0.2cm}

(4)
Finally, we check the results of the model. We assume that we start betting on the market at the time $t=0$ with $1000$ of monetary units and we stop when
we loose all of them.
In order to check the success of the model, we produce a simulation considering first that the reward function is purely obtained by the information of the market (Figure \ref{allocations}), and secondly using $50 \%$ of dreams.
To do this, we use the second part of the experience.
The system has been trained using all the information of the first $400$ minutes in the first case (Figure \ref{CUMSUMnodream}), and with just $50 \%$ of these states + $50 \%$ of dreams in the second one (Figure \ref{CUMSUMwithdream}). In these figures it can be seen
the value of the sum of the four products of the market at each state, where the investment that has been made in each of them has been the result of the application of the action/bet obtained in the previous stages.
The measure of the success of the models is given by the survival time.

For the first case (Figure \ref{CUMSUMnodream}) we have used the set of actions obtained for the set $M_0$, which was shown in Figure \ref{allocations}.
It is supposed that the situations should be similar than in the training part of the experience. However, in case the state $s$ was not exactly appearing in  the market situations that was recorded in the first part of the experience, we approximate its value by distance similarity applying the action $a_{s'},$ where $s'$ is the element of $M_0$ that satisfies that $d(s,s')$ attains its minimum.

The second figure (Figure \ref{CUMSUMwithdream}) shows the same cumulative result: the total value obtained at each state by applying to the same sequence of states the optimal sequence of actions, that has been obtained in  this case with a $50 \%$ of dreams. As the reader can see, the evolution and  the surveyance time are similar, and so the  success of both models is comparable. That is, the same result can be obtained by using the McShane extension of the $50 \%$ of known data instead of
$100 \%$ of real data.

\begin{remark}  \label{remrew}

\

\begin{itemize}

\item[$\circ$] Note that dreams and real states are of a different nature. Real states come directly from the observation of the system, while dreams are artificially generated states that mix real components---through some kind of interpolation---and also add some random components, as explained in this section.

\item[$\circ$] Our model allows to create an automatic forecasting system that introduce updated data at every moment. The direct implementation of our model in the financial markets provides an automatic system that advises the investor at all times on the best investment. 
Broadly speaking, applications for market data analysis would be given by an algorithm that would provide the analyst with a simulation of his investment, helping him to benefit from market trends.

\item[$\circ$] Unlike neural networks, the McShane formula provides a method that penalizes predicted reward values if the state under consideration is far from the states used to estimate its value (just take a look at the formula). It is an extrapolation based on continuity, which can be quantified thanks to the preservation of the Lipschitz constant in each state.

\item[$\circ$]
As we said in part 3 of Section \ref{S4}, other extension formulas could be used instead of the McShane extension. We can suggest an easy way to improve the choice: simply by choosing the best convex combination of the McShane and Whitney formulas (which in a way represent "extreme extensions", as suggested by the upper and lower labels used in \cite{aron}). The optimal parameter of the convex combination could be estimated using some real data from the problem and a Monte-Carlo estimate.  

\end{itemize}

\end{remark}

\section{Conclusions} \label{S5}

We have shown a reinforcement learning method to provide an expert system for investing in a financial market. The first introduced tool, that involves
approximation of a reward function by using metric similarity with other known states of the system, is based on a classic machine learning scheme on metric spaces
and the McShane extension of real functions preserving the Lipschitz constant.
Regarding this point, the main novelty is the non-standard metric that is used, that combines a geodesic distance  ---directly related with the cosine similarity of vectors and that models the directions of the trends
of the market---  and the Euclidean distance, which
cannot be defined as associated to a norm in the underlying finite dimensional linear space.

The second part of our technique consists on the development of a new reinforcement learning procedure that allows the use of a smaller set  $M_0$ of experiences on the financial market
to obtain a good investment tool to act in the market. Basically, we combine the use of approximation of the reward function on neighbors of $M_0$ with a Lipschitz-preserving
extension of the reward function by using the McShane formula. Thus, one of the contributions of the present paper is to show that  an expert system  for investment in  financial markets
can be done by substituting a great set of experiences on the particular markets by a reinforcement learning method based on the extension of Lipschitz   maps.
Since the results obtained are comparable, our technique opens up the possibility of building models of similar efficiency using much less data from experience.

\vspace{0.2cm}

\section{Acknowledgements}
This work was supported by   the Ministerio de Ciencia, Innovaci\'on y Universidades, Agencial Estatal de Investigaciones and FEDER (Spain) [grant number   MTM2016-77054-C2-1-P.]

\newpage

 \begin{figure}[h]
\vspace{0.1cm}
\centering
\includegraphics[width=0.4\textwidth]{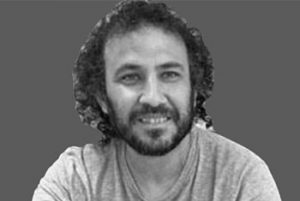}
\end{figure}

Jos\'e M. Calabuig is professor of Applied Mathematics at the Polytechnic University of Valencia and researcher at the Institute of Pure and Applied Mathematics of this university. His research work has focused on some Mathematical Analysis topics -mainly, space and the Banach Operator Theory-, in which he has published more than 40 research papers. He is also an active researcher in applications of mathematical structures in computing, economics and other applied areas, and is currently developing several research projects on anti-fraud computational techniques, amd applications of blockchain in public administration. This research paper shows some of the results on the theoretical aspects of these projects.

 \begin{figure}[h]
\vspace{0.01cm}
\centering
\includegraphics[width=0.4\textwidth]{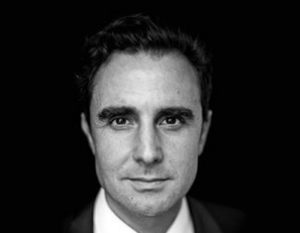}
\end{figure}

Herv\'e Falciani is a computer engineer and analyst who has developed an active career against economic fraud in the financial markets. He has collaborated on Internet security and tax fraud issues with several European governments, public administrations and social organizations. He is currently participating in several anti-fraud projects as an analyst and researcher as a collaborator of the Institute of Pure and Applied Mathematics of the Polytechnic University of Valencia.
He also worked for some years as a Data Scientist at INRIA, developing some deep learning tools for the detection of anomalous events, mainly in terms of graphical data.

\newpage

 \begin{figure}[h]
\vspace{0.01cm}
\centering
\includegraphics[width=0.4\textwidth]{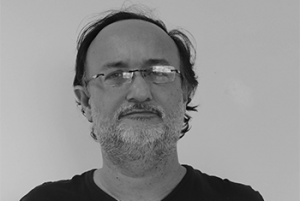}
\end{figure}

Enrique A. S\'anchez P\'erez is also professor of Applied Mathematics at the Polytechnic University of Valencia and researcher at the Institute of Pure and Applied Mathematics.
He is an active researcher in Functional Analysis, in which he has published more than 130 articles and several books. He has also worked on interdisciplinary projects with research groups in other sciences, such as Health Sciences, Acoustics, Information Science and Computer Science. He also participates in the project for the development of the mathematical bases of  anti-fraud computer tools.

\end{document}